# Magnetic ground state and strain-mediated chiral-like atomic distortions behavior in two-dimensional rectangular spin lattice

Yu Liao,[1,#] Yueqiao Qu,[2,#] Zuo Li,[1] Yu Chen,[3] Liang Liu,[4] Jun-Zhong Wang,[1]* and Gang Yao[1]*

[1]Chongqing key Laboratory of Micro & Nano Structure Optoelectronics, School of Physical Science and Technology, Southwest University, Chongqing 400715, China
[2] Key Laboratory of Artificial Structures and Quantum Control (Ministry of Education), School of Physics and Astronomy, Shanghai Jiao Tong University, Shanghai 200240, China
[3]School of Science, Inner Mongolia University of Technology, Hohhot 010051, China
[4]School of Physics, State Key Laboratory for Crystal Materials, Shandong University, Jinan 250100, China

[#]These authors contributed equally: Yu Liao, Yueqiao Qu
*Corresponding authors. Emails: jzwangcn@swu.edu.cn, yaogang@swu.edu.cn

**Due to the large perpendicular magnetic anisotropy originating from spin-orbit coupling, magnetoelastic coupling is generally reported in easy-plane magnets with rectangular lattice where the easy magnetization is coupled with the lattice direction, while the acquisition of a novel coupling, beyond the easy-plane ferromagnets, in two-dimensional (2D) materials remains unknown. Here, by employing the density functional theory calculations, we demonstrate this feasibility with the discovery of long-range ferromagnetic ordering and elastic strain-mediated chiral-like atomic distortions behavior in a newly tetragonal As-Fe-As trilayer (t-FeAs monolayer), which shows large perpendicular magnetic anisotropy, robust ferromagnetic ordering, and in-plane ferroelasticity. We firstly point out that obvious limits exist when using the four magnetic configurations to determine the magnetic ground state for a rectangular spin lattice even if more exchange interaction parameters are included. A four-state mapping analysis is carefully examined for t-FeAs, where the calculated Curie temperature, $T_c$, is 435 K, which is higher than most reported 2D magnets, and can be further tuned by appropriate strains. Intriguingly, the chiral-like atomic distortion behavior of the Fe sub-layer is scanning tunneling microscopy characterizable, which can switch the magnetization axis between the out-of-plane and in-plane direction. This unusual finding of ferroelastic manipulation of both the atomic displacement and spin properties makes t-FeAs a promising candidate for future spintronics and also provides the possibility for exploring unprecedented coupling physics.**

# I. INTRODUCTION

Since the essential elements within information devices are consistently shrinking to the nanometer scale as well as the initial discovery of graphene, 2D magnetic materials with high tunable physical properties have attracted particular research interests [1-5]. Multiferroicity, an intriguing physical property arising from the combination of more than two ferroic orders among ferromagnetism (FM), ferroelasticity (FEL), ferroelecticity, and ferrotroidicity, holds a great potential for multiple-state memory, switches, and computing [6-9]. As an example, a magetoelectric material allows for the control of magnetism (electric polarization) by external electric field (magnetic field). Comparatively, FM-FEL materials allow the operation through the mode of "writing mechanically and reading magnetically", and also promise an extension of the spectrum of applications for multiferroic materials. This material is particularly advantageous as it opens the door for realizing the combination of nonvolatile memory and FM in a single-material circuit. However, ferroelectricity commonly necessitates vacant $d/f$ orbitals, whereas FM is often linked to partially filled $d/f$ orbitals. The conflicting origins of these ferroic orders makes the inherent magetoelectric materials very scarce. On the simplest level, due to the crystal anisotropy, a 2D ferromagnet with rectangular lattice is inherently FM-FEL multiferroic. Unfortunately, although FEL has been experimental demonstrated or theoretical predicated in 2D systems, such as layered-perovskite thin films, transition metal dichalcogenides, phosphorene and phosphorene analogues, as well as honeycomb lattice (including graphene, BN, stanene, etc.) [10-16], their nonmagnetic nature greatly restrict their applications in spintronics. To this end, 2D ferroelastic lattices with desirable magnetic and mechanical properties for scalable spintronic device applications are in urgent need.

The absence of magnetism in many 2D materials have motivated great efforts to artificially introduce or control spin ordering via doping [17,18], defect engineering [19,20], proximity coupling [21,22]. Typically, arising from the van Hove singularity of valence band edges, FM is derived in nonmagnetic but ferroelastic monolayers of $α$-SnO and $α$-PbO under carrier doping [17,18]. So far, plenty of 2D materials with coexisting FEL and intrinsic FM characters have also been reported [23-28], but few of them possess both high-temperature magnetism and observable FEL switching signals. And especially, there are even less materials that exhibit coupling between these two orders, which have severely hampered their applications for functional devices. For instance, in 2D ferromagnets with rectangular lattice, FEL is inherent. This category of materials therefore distinguishes themselves among 2D crystals and could potentially serve as the highly desired monolayer magnetic component which is essential for creating comprehensive 2D spintronic devices, offering the additional benefit of enabling ferroelastic adjustment of the magnetic state. Nevertheless, creating novel 2D FM/FEL crystals possessing room-temperature intrinsic FM, low ferroelastic barriers, strong switching signals, and importantly the cross coupling between these orders remains imperative and exceedingly arduous.

Here, via first-principles calculations, molecular dynamics, and Monte Carlo simulations, we unveil that monolayer tetragonal FeAs (t-FeAs) with rectangular lattice, a brand-new phase of the reported 2D FeAs family, exhibits an interesting coexistence of ferromagnetic ordering and multiferroic couplings. We estimate the critical temperature for the magnetic transition, $T_c$, and suggest that mechanical strains can switch the $T_c$. Owing to the inherent crystal anisotropy, t-FeAs exhibits a strain-driven 90º variant switching with moderate barrier and strong transition signal, making it FM-FEL multiferroic. Furthermore, a chiral-like atomic displacement is observed for the first time, followed by a magnetization switchable behavior. Our work indicates that t-FeAs can be an excellent candidate for advanced device applications.

**II. COMPUTATIONAL PROCEDURES**
Our first-principles calculations are carried out within spin-polarized density functional theory (DFT) by using Vienna *ab initio* simulation package (VASP) [29,30]. The generalized gradient approximation (GGA) of the Perdew-Burke-Ernzerhof (PBE) [31] functional for the exchange and correlation potential is adopted. The electron-ion interactions are described using the projector augmented wave (PAW) method [29]. The localized Fe-*d* orbitals are treated employing the GGA + $U$ approach with a Hubbard $U_{eff} = U - J = 3$ eV. To ensure the decoupling between the monolayer and its periodic layer, a vacuum space of 20 Å is added along the *c* axis. We find it is sufficient to take a combination of cutoff energy of 500 eV for plane-wave basis expansion and a $\Gamma$-centered 15 × 15 × 1 Monkhorst-Pack grid for the Brillouin zone integration to obtain the convergence which is $10^{-6}$ eV for the electronic self-consistence loop and $10^{-2}$ eV/Å$^2$ for the Hellmann-Feynman forces on each atom. The magnetocrystalline anisotropy energies (MAEs) are calculated using a dense *k* mesh of 31 × 31 × 1, and the spin-orbit coupling (SOC) effect is included. The nudged elastic band (NEB) method is also implemented to estimate the energy barriers between different phases [32]. The phonon dispersions are calculated with the finite displacement method and the PBE functional on a 3 × 3 × 1 supercell using the Phonopy code [33]. A*b initio* molecular dynamics (AIMD) simulations are performed at 300 K using NVT ensemble up to 10 ps with a time step of 2.5 fs. The scanning tunneling microscopy (STM) simulations are performed with the Tersoff-Hamann method at constant height mode [34]. The distance between STM tip and the sample surface is set at 3 Å. The geometries and spin charge densities are visualized by VESTA package [35].

**III. Crystal structure and stability of t-FeAs**
The experimentally synthesized single crystal TMFe$_2$As$_2$ (TM = Ba, Ca, and Cs) bulk crystallize with space group of *I*4*/mmm* and consist of alternating As-Fe-As trilayer and TM atom layer stacked along the *c* axis, as displayed in Fig. 1(a). The monolayer FeAs [Fig. 1(b)], which is the focus of this work, has a tetragonal lattice in the *P*4/*nmm* space group and is isostructural to the famous interface high-temperature superconductor of FeSe layer [36,37]. By modeling the exfoliation process, the exfoliation energies (0.103 ~ 0.181 eV/Å$^2$; see Fig. S1 in Supporting Information [38]) are comparable to those of MXenes and 2D non-vdw materials (0.086 ~ 0.205 eV/Å$^2$) [39-42]. Thus, FeAs nanosheet could be obtained from its bulk counterparts. Nevertheless, after successful optimizations, monolayer FeAs retains its tetragonal puckered form, but changes from a square lattice into a rectangular one, with lattice constants of $a = 2.83$ Å and $b = 3.97$ Å, and the point symmetry thus changes from $C_{4v}$ into $C_{2v}$ (Fig. 1c).

We note that our optimized lattice is considerably similar to a recent proposed phase, which also hosts a rectangular cell but moderate aspect ratio ($a = 3.17$ Å and $b = 3.80$ Å), denoted as MR-FeAs [43]. The dynamical and thermal stability of both systems are evidenced by the phonon spectra and AIMD simulations (Fig. S2). Besides, three different phases of FeAs monolayer were also predicated previously [44], including two forms of square lattice (FeAs-I and FeAs-III) and a trigonal structure one (FeAs-II). As compared in Fig. S3, our proposed 2D lattice of FeAs is only 117 meV/atom higher than that of the most stable FeAs-I phase, and much lower than the remaining three phases in energy of at least 33 meV/atom, indicating its competitive chemical stability. As a key factor in theoretically designed 2D materials, the good kinetic stability is supported by a large positive cohesive energy ($E_{coh}$). For t-FeAs, $E_{coh} = (2 \times E_{Fe} + 2 \times E_{As} - E_{FeAs})/4$, where $E_{Fe}$, $E_{As}$, and $E_{FeAs}$ are the total energies of Fe, As atom, and 2D FeAs unitcell, respectively. We obtain $E_{coh} = 2.89$ eV/atom, which is comparable with other well-known materials, such as the experimental fabricated planar hexa-coordinated $Cu_2Si$ (3.46 eV/atom) [45] and $Cu_2Ge$ (3.17 eV/atom) [46]. These above results support the thermodynamic feasibility of synthesizing the t-FeAs monolayer in experiments.

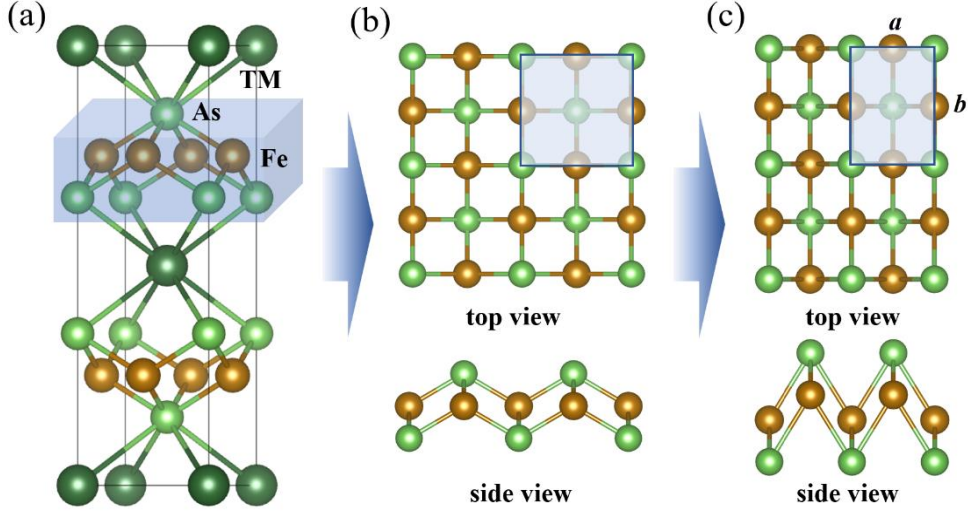

FIG. 1. Geometric structures of monolayer FeAs and its bulk counterpart. (a) Crystal structure shared by the $TMFe_2As_2$ (TM = Ba, Ca, and Cs) bulk phases, where one sub-layer of FeAs in its unit cell is denoted by the blue shaded cuboid. The TM, Fe, and As atoms are displayed in dark green, orange, and green, respectively. The black box shows the unit cell. (b, c) The top and side views of single-layer FeAs before and after optimization, respectively. The bule shaded regions denote the primitive unit cells.

## IV. PROCEDURE TO DETERMINE THE HIGH CURIE TEMPERATURE OF t-FeAs
### A. Limitations of four spin configurations method

After having established the structure and stability of the new 2D FeAs, its magnetic and electronic properties are further investigated. Generally, for a 2D magnet, one FM and several AFM configurations are considered to reveal the preferred spin ordering by comparing their total energies. Then the spin-exchange interactions can be extracted by mapping the DFT energies to the Heisenberg spin Hamiltonian:

$$H = \sum_{(i,j)} J_1 \vec{S_i} \cdot \vec{S_j} + \sum_{(i,j)} J_2 \vec{S_i} \cdot \vec{S_j} + \sum_{(i,j)} J_3 \vec{S_i} \cdot \vec{S_j} + A(S_i^z)^2, \qquad (1)$$

in Eq. (1), $S_i$ ($S_j$) is the unit vector ($|S| = 1$) of local spin at the $i$th ($j$th) Fe atom. $J_{1,2,3}$ is the isotropic Heisenberg exchange coupling between two different Fe atoms at sites $i$ and $j$ [Fig. 2(a)]. Negative (positive) $J_i$ implies FM (AFM) exchange interaction. $A$ is the single-site magnetic anisotropy parameter, and $S_i^z$ represents $z$ (*i.e.*, out-of-plane orientation) component of $S$. As displayed in Fig. 2(a-e), five possible magnetic configurations are adopted for t-FeAs, and their total energies are compared in Fig. 2(f). The energy of FM configuration is found much lower than that of $AFM_1$, $AFM_2$, and $AFM_3$, but slightly higher in energy than $AFM_4$ by 10.4 meV/unit cell. What's noteworthy is that a FM ground state will be extracted if considering only FM and AFM$_1$ orders, because of the resulted large negative $J_1 = (E_{FM} - E_{AFM1})/32 = -39.07$ meV/Fe. A similar puzzle was previously reported in 2D monolayers of CrSiSe$_3$ [47] and strained Janus Cr$_2$XYTe$_6$ (X, Y = Si, Ge, Sn, and X ≠ Y) [48], for which ignoring $J_2$ and $J_3$ had yielded an incorrect magnetic ground state. It is thus necessary to study the preferred magnetic coupling by performing MC simulation with multiple exchange interactions, usually up to the third one ($J_3$), into consideration.

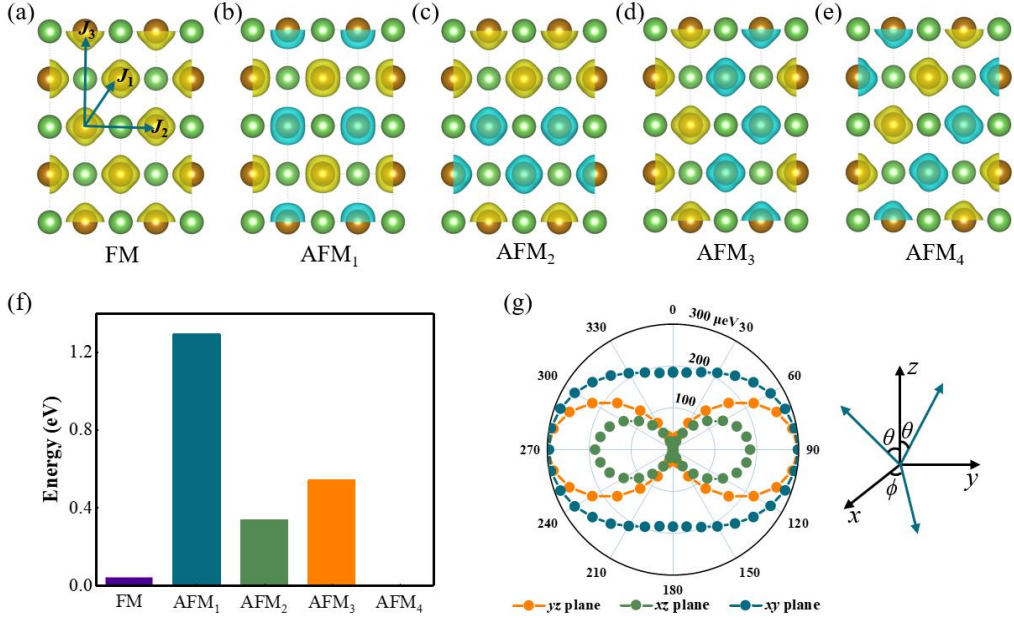

FIG. 2. Magnetic properties. (a-e) The isosurface plots (0.67 e/Å$^3$) of the spatial spin density for t-FeAs monolayer in FM, and four possible antiferromagnetic configurations: AFM$_1$, AFM$_2$, AFM$_3$, and AFM$_4$. The majority and minority spin charges are indicated by isosurfaces with yellow and blue colors, respectively. Bule arrows represent the exchange path of nearest-neighbor (NN), second-, and third-NN Heisenberg exchange interactions ($J_1$, $J_2$, and $J_3$). (f) Relative total energies of the various magnetic configurations shown in (a-e), where the total energy of the AFM$_4$ is taken as a reference. The energies are calculated without the inclusion of spin-orbit coupling. (g) Angular dependence of MAE, obtained by rotating the spin orientations within $xy$, $xz$, and $yz$ planes. Polar angle $\theta = 0°$ pointing along the positive $z$ axis in the $xz/yz$ planes, while along the $x$ axis in the $xy$ plane, as indicated.

For the different spin configurations [Fig. 2(a)-(e)], the total energy can be expressed as:
$$E_{FM}/4 = E_0 + (4J_1 + 2J_2 + 2J_3)|S|^2 - A|S|^2, \qquad (2)$$
$$E_{AFM1}/4 = E_0 + (-4J_1 + 2J_2 + 2J_3)|S|^2 - A|S|^2, \qquad (3)$$
$$E_{AFM2}/4 = E_0 + (0J_1 + 2J_2 - 2J_3)|S|^2 - A|S|^2, \qquad (4)$$
$$E_{AFM3}/4 = E_0 + (0J_1 - 2J_2 + 2J_3)|S|^2 - A|S|^2, \qquad (5)$$
$$E_{AFM4}/4 = E_0 + (0J_1 - 2J_2 - 2J_3)|S|^2 - A|S|^2, \qquad (6)$$

where $E_0$ is the total energy independent to different magnetic states. Note that we set $|S| = 1$, and parameter $A$ in Eq. (1) is kept. As listed in Table I, a relatively large negative $J_1$ originated from the direct exchange interaction is always extracted. In addition, $J_1$, $J_2$, and $J_3$ are much dependent on the choice of spin configurations. Similar result was previously reported for monolayer $Cr_2Te_6$ [49]. By performing MC simulations based on the spin Hamiltonian parametrized with the MAEs and $J_{i,j}$ in Eq. (1), nonzero $T_c$ are always obtained (Table I), implying FM coupling in t-FeAs, which confirms again the primary responsibility of $J_1$ for the magnetic ground state. In addition, the $T_c$'s are significant different, with values ranging from 13 to 160 K. Therefore, it can be concluded that the four magnetic configurations method has obvious limit in determining exchange interaction parameters, and hence the magnetic transition temperature for a rectangular spin lattice.

In the end of this section, we examine the MAE, which defined as MAE = $E_\parallel$ - $E_\perp$, where $E_\parallel$ and $E_\perp$ are the total energies of in-plane, $M_a$[100] and $M_b$[010], and out-of-plane $M_z$[001] spin configurations, respectively. According to the Mermin-Wagner theorem [50], the magnetically ordered phase of 2D materials is dependent on the MAE and the distance dependence of the exchange interaction. Also, in a 2D system, a stable long-range magnetic ordering which is robust against the thermal fluctuations is beneficial to magnetic data storage. Figure 2(g) displays the MAE as a function of polar angles $\theta$ on the $xy$, $xz$, and $yz$ planes. The lowest energy state corresponds to the [001] direction, indicating out-of-plane magnetization. Using the total energy corresponding to the $M_z$[001] direction as a reference, we obtain $MAE_{[100]-[001]}$ = 184 μeV/Fe atom and $MAE_{[010]-[001]}$ = 299 μeV/Fe, suggesting large out-of-plane magnetic anisotropy. Therefore, t-FeAs belongs to the category of 2D Ising magnets possessing a long-range (anti)ferromagnetically ordered low temperature phase, verifying the reasonableness of the spin Hamiltonian we selected. Note that the MAE value for t-FeAs is more than 3 times larger than that of cubic Ni (2.7 μeV/atom), and Fe (1.4 μeV/atom), and Co (65 μeV/atom) [51,52]. The large MAE values render t-FeAs appropriate for applications in magnetoelectronics.

TABLE I. The spin configuration results for the Heisenberg exchange parameters extracted from Eqs. (2)-(6) without SOC, and the estimated Curie temperature ($T_c$) of t-FeAs.

| Configurations | $J_1$ | $J_2$ | $J_3$ | $T_c$ |
| --- | --- | --- | --- | --- |
| FM+AFM$_{1,2,3}$ | -39.07 | 10.50 | 20.71 | 150 |
| FM+AFM$_{1,2,4}$ | -39.07 | 27.94 | 20.71 | 13 |
| FM+AFM$_{1,3,4}$ | -39.07 | 10.50 | 22.53 | 160 |
| FM+AFM$_{2,3,4}$ | -52.15 | 27.94 | 33.79 | 16 |

**B. Estimate of $T_c$ based on the four-state mapping analysis**

To make the exchange parameters, and hence the $T_c$ clear, a four-state mapping analysis is applied [53,54]. In this method, the $J$ related total energy of magnetic configuration is written as

$$E_{\text{spin}} = J_{12}\vec{S_1}\cdot\vec{S_2} + \vec{S_1}\cdot\vec{K_1} + \vec{S_2}\cdot\vec{K_2} + E_{\text{other}}, \quad (2)$$

where $\vec{K_1} = \sum_{i\neq 1,2} J_{1i}\vec{S_i}$, $\vec{K_2} = \sum_{i\neq 1,2} J_{2i}\vec{S_i}$, and $E_{\text{other}} = \sum_{i\neq 1,2} J_{i,j}\vec{S_i}\cdot\vec{S_j}$. $J_{12}$ denotes the exchange interaction between two spins at sites 1 and 2. Four collinear spin states including (1) $|S|_1^z = |S|$, $|S|_2^z = |S|$; (2) $|S|_1^z = |S|$, $|S|_2^z = -|S|$; (3) $|S|_1^z = -|S|$, $|S|_2^z = |S|$; and (4) $|S|_1^z = -|S|$, $|S|_2^z = -|S|$ are considered. Meanwhile, the spin orientations of other spin sites are always set to the same. This method is irrelevant to the spin configurations shown in Fig. 2(a-e), and a more accurate $J$ value will be extracted if a larger spin lattice is used.

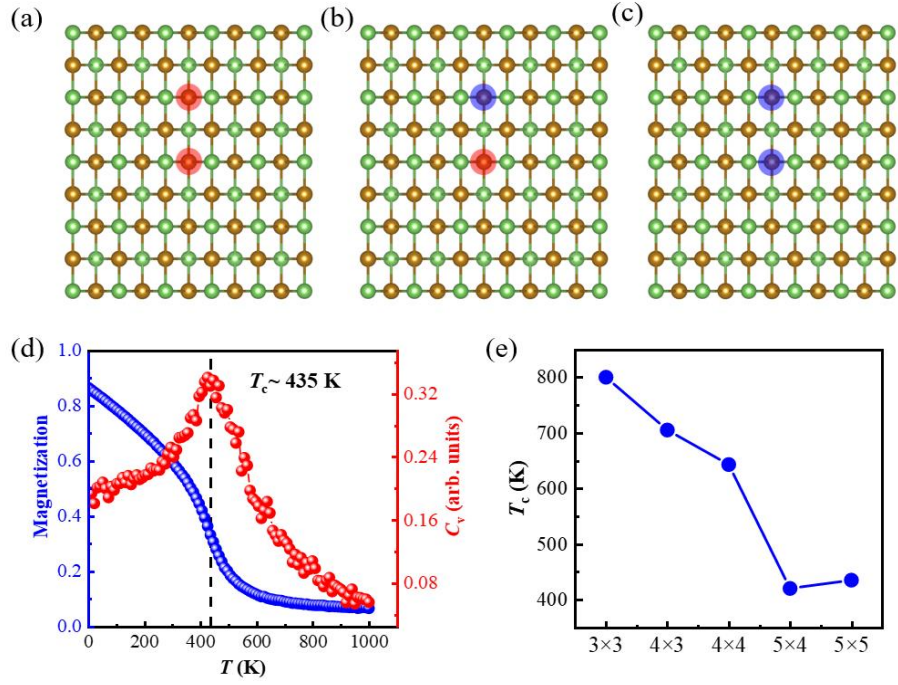

FIG 3. Configurations with different ordered spin states and MC simulations. (a) ↑↑ spin ordering, (b) ↑↓ or ↓↑ spin orderings, and (c) ↓↓ spin ordering for estimating $J_3$. The up-spins and down-spins on the Fe atoms are denoted as red and blue shaded cycles, respectively. (d) Magnetization and specific heat, $C_v$, with respective to temperature. (e) Effect of the supercell size on the $T_c$.

Take $J_3$ calculation as an example, the corresponding spin states based on a $5 \times 5 \times 1$ supercell are displayed in Fig. 3(a-c). The energy expressions for these four spin states are

$$E_1 = E_0 + E_{\text{other}} + J_3|\mathbf{S}|^2 + K_1|\mathbf{S}| + K_2|\mathbf{S}|, \tag{3}$$
$$E_2 = E_0 + E_{\text{other}} - J_3|\mathbf{S}|^2 + K_1|\mathbf{S}| - K_2|\mathbf{S}|, \tag{4}$$
$$E_3 = E_0 + E_{\text{other}} - J_3|\mathbf{S}|^2 - K_1|\mathbf{S}| + K_2|\mathbf{S}|, \tag{5}$$
$$E_4 = E_0 + E_{\text{other}} + J_3|\mathbf{S}|^2 - K_1|\mathbf{S}| - K_2|\mathbf{S}|, \tag{6}$$

and $J_3$ can be deduced from Eqs. (3)-(6):

$$J_3 = \frac{E_1 - E_2 - E_3 + E_4}{4|\mathbf{S}|^2}, \tag{7}$$

we obtain $J_1 = -44.06$ meV, $J_2 = -16.64$ meV, and $J_3 = 2.08$ meV. The NN exchange parameter $J_1$, originated from the direct-exchange interaction, is much larger than $J_2$ and $J_3$, derived from the super-exchange interactions. This means that the direct-exchange interaction contributes dominantly to the FM arrangement of t-FeAs, constant with the results of the four spin configurations method.

From the simulated magnetization and specific heat curves in Fig. 3(d), the $T_c$ of monolayer t-FeAs is found to be 435 K. The convergence of $T_c$ is also tested with different supercell sizes, as shown in Fig. 3(e). The $T_c$ difference between supercell of $5 \times 4 \times 1$ and $5 \times 5 \times 1$ is only 15 K. When considering neighboring exchange interactions up to $J_5$, $J_4 = 1.46$ meV and $J_5 = 0.46$ meV, a very close value ($T_c = 410$ K, see Fig. S4) is obtained. Therefore, the calculated $5 \times 5 \times 1$ supercell and numbers of spin-exchange interactions are enough to achieve the desired computation accuracy. For comparison, the $T_c$ value of the intrinsic FM semiconductor CrI$_3$ estimated by the same method is 57 K for $U_{eff} = 0$ eV and 78 K for $U_{eff}$ = 3 eV (Fig. S5), very close to the experimental measurement (∼ 45 K) [2] and previous theoretical values (35 K ∼ 75 K) [55,56], confirming the validity of our method.

Collectively, we demonstrated the importance of either including more exchange interaction parameters in determining the magnetic ground state or to employ the four-state mapping analysis for the calculation of the $T_c$ of 2D magnets with rectangular spin lattice. The underlying mechanism of FM can be disclosed in more details from its electronic structures. The calculated electronic band structure reveals that the t-FeAs monolayer is a spin-polarized metal [Fig. S6(a)]. From the spin charge density plotted in Fig. 2a, the major magnetism of t-FeAs is contributed by spin-polarized 3$d$ orbitals of Fe atoms. The total spin magnetic moment is ∼3.0 $\mu_B$ per Fe atom, indicating a large spin polarization in this 2D crystal. Further, the local magnetic moment on Fe and As atoms are about 3.048 $\mu_B$ and -0.058 $\mu_B$, respectively. As such, the Fe atoms are antiferromagnetic coupling to their neighboring As atoms. This behavior, along with the metallic feature, suggests that the carrier-mediated double-exchange mechanism is responsible for the ferromagnetic state of t-FeAs monolayer. On the basis of the charge transfer and crystal field theory, the local magnetic moment of Fe can be understood. Because of the tetrahedron crystal field of Fe atoms, the $d$ orbitals split into a set of upper triple-degenerated $t_{2g}$ ($d_{xy}$, $d_{x^2-y^2}$, $d_{xz}$) and lower double-degenerated $e_g$ ($d_{yz}$, $d_z^2$), which further split into five nondegenerate orbitals, as schemed in Fig. S6(b), consistent with the calculated orbital-resolved DOS [Fig. S6(c)]. In the minority spin channel, all the five orbitals are occupied, while only $d_z^2$ and $d_{xz}$ in the majority ones are occupied and the other three orbitals empty. Thus, a net magnetic moment of ∼3.0 $\mu_B$ is observed, agreeing with the result from our DFT calculations.

## V. MULTIFERROIC COUPLINGS
### A. Ferroelastic strain-induced chiral-like atomic distortions behavior

Given the rectangular lattice of t-FeAs at its ground state, the in-plane elastic switching with a non-volatile strain is inherent. Figure 4(a) illustrates the general ferroelastic switching pathway for a 2D tetragonal material [17,18,25], where the structure of the initial/FEL-I state is the same as that of the final (*i.e.*, FEL-II) one after a rotation of 90º. These two FEL states are connected by an intermediate state with a square lattice, *i.e.*, paraelastic (PA) state. The calculated elastic energy $\Delta E_{el}$ for t-FeAs versus strain varying $\varepsilon_{xx}$ and $\varepsilon_{yy}$ for lattice constants in the *x* and *y* directions, and the internal positions are fully relaxed. The energy profile from initial state to final state shown in Fig. 4(b) has two degenerate minima at $\varepsilon_{xx}$ = 0% ($\varepsilon_{yy}$ = 0%) and $\varepsilon_{xx}$ = 40.3% ($\varepsilon_{yy}$ = -28.7%). Clearly, t-FeAs do exhibit intrinsic FEL. Therefore, 2D t-FeAs crystal is multiferroic that hosts FM and FEL simultaneously. It should be emphasized that in contrast to the centrosymmetric one, a non-centrosymmetric phase named twist phase hereafter is observed at PA state [Fig. 4(c)], which indeed lower in energy than the centrosymmetric. Quite interestingly, in this 2D planar, the Fe sublattice features a chiral-like lattice distortions as indicated, while the four As atoms in internal positions of (1/4, 1/4, *z*), (1/4, 3/4, *z*), (3/4, 1/4, *z*), and (3/4, 3/4, *z*) are equally moving toward the corner of the cell, leading to a $C_{2v}$ symmetry rather a $C_{4v}$ one. The energy- $\varepsilon_{xx}$ curve for this phase is then checked. The lattice constants *a* and *b*, during this process, are fixed at a given strain, while the internal coordinates are relaxed. A similar FEL trend is observed (Fig. 4b). Compared with the normal phase, the twist phase possesses higher energy at FEL-I state but is energetically stable at PA state by 10 meV/atom and 60 meV/atom, respectively. The tiny energy difference between these two phases at FEL-I state indicates the metastable nature of the twist phase. The critical point for the structural phase transition occurs at about $\varepsilon_{xx} = 6\%$ ($\varepsilon_{yy} = -4.3\%$), which is feasible in experiment as evidenced by the stress-strain curves (Fig. S7). It should be noted that similar results are obtained on a larger supercell.

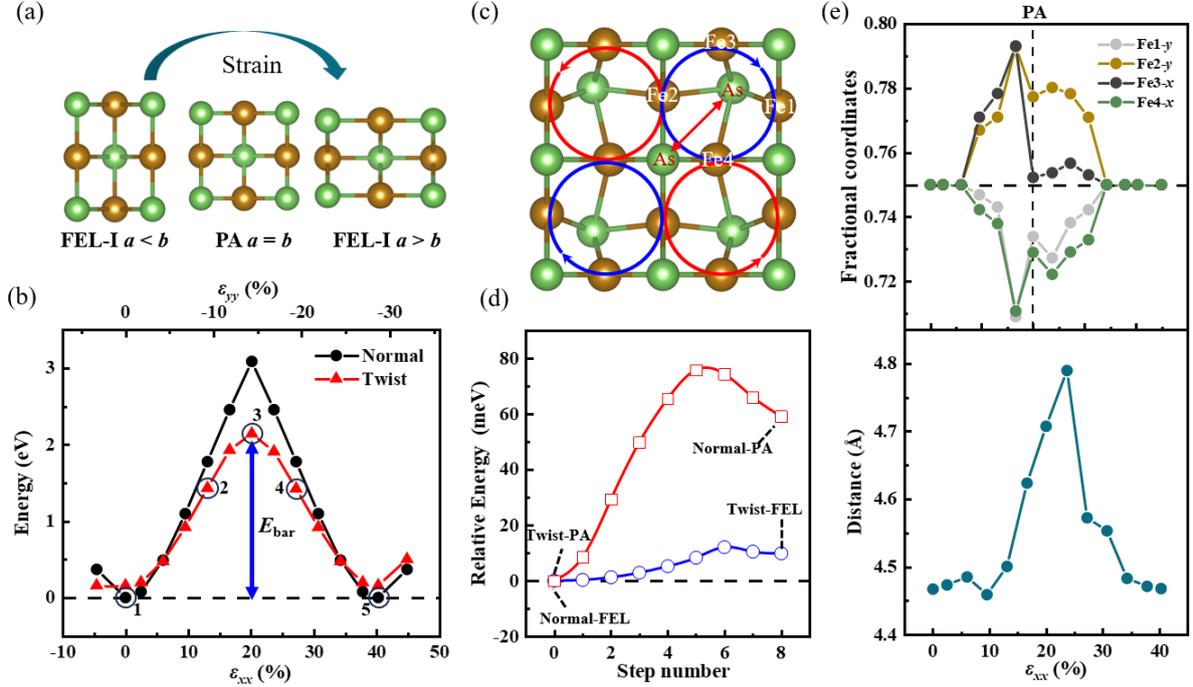

FIG. 4. In-plane FEL of t-FeAs. (a) General ferroelastic switching pathway of 2D tetragonal materials. (b) Ferroelastic switching energy profile of t-FeAs. The energy of the initial state (FEL-I) is set to zero. (c) The optimized structure of the twist FeAs at PA state, in which the deviation directions of the Fe and As atoms are indicated. (d) Changes of relative energies per atom during the phase transition between normal and twist phases for FEL-I and PA states. (e) Variation of the Fe-displacements (upper panel) and As-As distance (lower panel) with the FEL strain.

The phase transition between these two phases at FEL-I and PA states are studied by gradually moving the internal positions [Fig. 4(d)]. One can see that, although the FEL-I state of the twist phase is just a little higher in energy than that of the normal phase, there is a small energy barrier (0.44 meV per Fe atom) between them, indicating that the metastable twist phase at FEL-I state can survive at a high temperature without transforming to the normal one. To characterize this tendency more visibly, the offsets of the internal coordinates are summarized in Fig. 4(e) as a function of elastic strain. Note that when the elastic strain deviates from the PA state, the deviations on both sides are asymmetric. Importantly, in the range from $\varepsilon_{xx} = 6$ to 16.6%, the deviation of the Fe atoms from their balance positions increases as elastic strain increases linearly. The large monotonically increased tendency indicates the close relationship between the atomic displacements and ferroelastic strain, which is expected for establishing the sensor to identify the coupling between these two orders.

To further judge the feasibility and robustness of FEL in t-FeAs. The overall transition barrier associated with the ferroelastic lattice rotation is calculated to be 134 meV/atom, which is lower than that of phosphorene (200 meV/atom) [57] and BP$_5$ (320 meV/atom) [58], but very close to that of α-MPI (M = Zr, Hf) (~135 meV/atom) [59]. Generally, a moderate energy barrier is desired in experiments, for the reason that a small one indicates the FEL state to be unstable, while a high one makes the ferroelastic lattice rotation achieve hardly. Therefore, the moderate value for t-FeAs renders the switchable anisotropic properties in this system upon the external stress accessible experimentally. Besides, another key factor for the FEL is the reversible strain defined as $(|b|/|a| - 1) \times 100\%$, which characterizes the structural anisotropy. The reversible ferroelastic strain for t-FeAs is calculated to be 40.3%. Compared with previously proposed ferroelastic materials, this value is much larger than those of VP and VAs monolayers (0.11% and 1.52%) [27], and comparable to that of phosphorene (37.9%) [57], $BP_5$ (41.4%) [58], and Cr-based Janus monolayers (22% ~ 36.5%) [26]. Therefore, t-FeAs would possess strong switching signal and obvious structural anisotropic differences.

**B. Coupling between atomic deviation and magnetization**

It is well known that the absolute positions of atoms in the non-surface-layer are relatively difficult to measure. In order to obtain deeper understanding of the electronic structure and surface morphology, and to aid in upcoming experimental characterization efforts, the scanning tunneling microscope (STM) images of t-FeAs monolayer are simulated in the entire FEL switching process. Enumerated in Fig. 5(a) are a series of structural models and their corresponding simulated STM topographies of t-FeAs monolayer as the FEL strain is gradually increased. The bright signals correspond to the topmost As atoms. Obviously, the intensity distributions of the topmost As-layers are different, indicating that the electron states in the vicinity of the Fermi level for the normal and twist phases are significantly different.

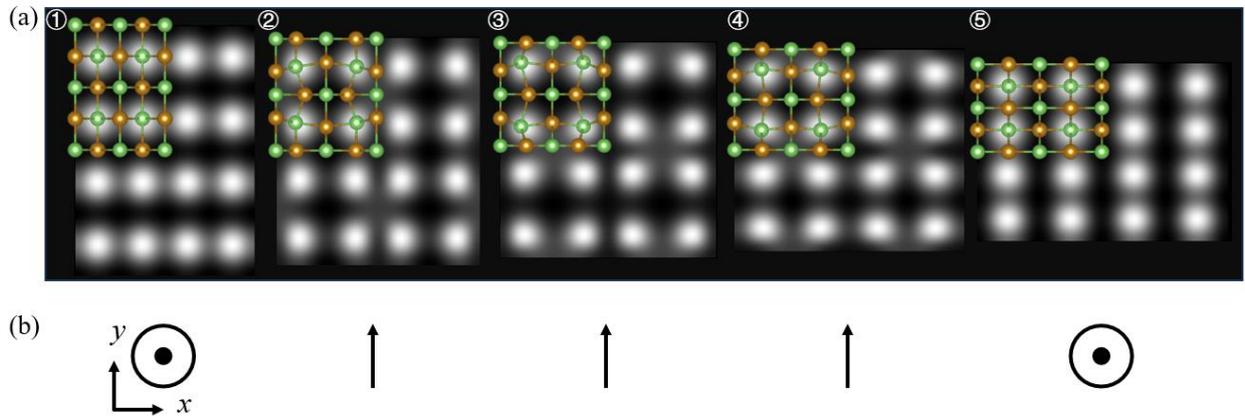

FIG. 5. Multiple multiferroic couplings. (a) The structural models and simulated STM images of t-FeAs under various ferroelastic strains, as indicated by cycles in Fig. 4(b). The STM images are shown under a bias voltage of +0.1 V. (b) The corresponding easy axis for (a).

One central issue that naturally arises is whether a possible magnetoelastic coupling followed by the structural phase transition exists. In this regard, we investigate the intercoupling between the ferroelastic strain and MAE. The detailed evaluation of MAE in the whole range of ferroelastic switching is displayed in Fig. S8. With a small strain applied, t-FeAs retains out-of-plane magnetization. Further increase the strain, the easy axis changes from out-of-plane direction into in-plane $b$ direction at $\varepsilon_{xx} = 13.0\%$, and can be further returned into the out-of-plane one at ~31%, leading to an interesting correlation between the easy axis and FEL strain. The tunable easy axis can be explained by the Fe-*d* orbitals appearing near the Fermin level, according to the second order perturbation theory [60-62], which allows for the effective management of spin injection and detection. The preferred spin orientations corresponding to these STM images are shown in Fig. 5(b). In addition to the magnetic atomic chiral-like deviation behavior, the spin properties are also significantly linked with in-plane elastic strain, offering strong multiferroic couplings in t-FeAs. Therefore, it is possible that both the spin properties and the chiral-like atomic distortions of the Fe-layer can be characterized or identified by the STM images through the changes of the intensity distribution of the surface As atoms. We expect that these features of 2D t-FeAs monolayer provide more information for identifying the newly discovered chiral-like atomic distortion behaviors, and also accelerate the possibility of its utilization in the near future.

## VI. DISSCUSSION
### A. Tensile strain switched ferromagnetism

Now, in-plane tensile strain is considered to manipulate the magnetic properties in t-FeAs. Previous works indicated that the MAE exhibits remarkable strain dependence in monolayer $VI_3$ [63], $CrX_3$ (X = Cl, Br, and I) [64], and $CrWGe_2Te_6$ [65]. Strain-tunable magnetism was also experimentally demonstrated at oxide domain walls [66]. Figure 6 shows the variation in $T_c$ and MAE in t-FeAs with tensile strain. The strain can be expressed as $\varepsilon = (l - l_0)/l_0 \times 100\%$, where $l_0$ and $l$ represent the lattice constants without and with the strain, respectively. The strain is uniformly applied along both *a* and *b* directions, or only along *a*/*b* direction, as shown in the insets of Fig. 6(a)-(c). Here, we consider the tensile strain up to 6%, which may be experimentally accessible, as evidenced by the stress-strain curves (Fig. S7).

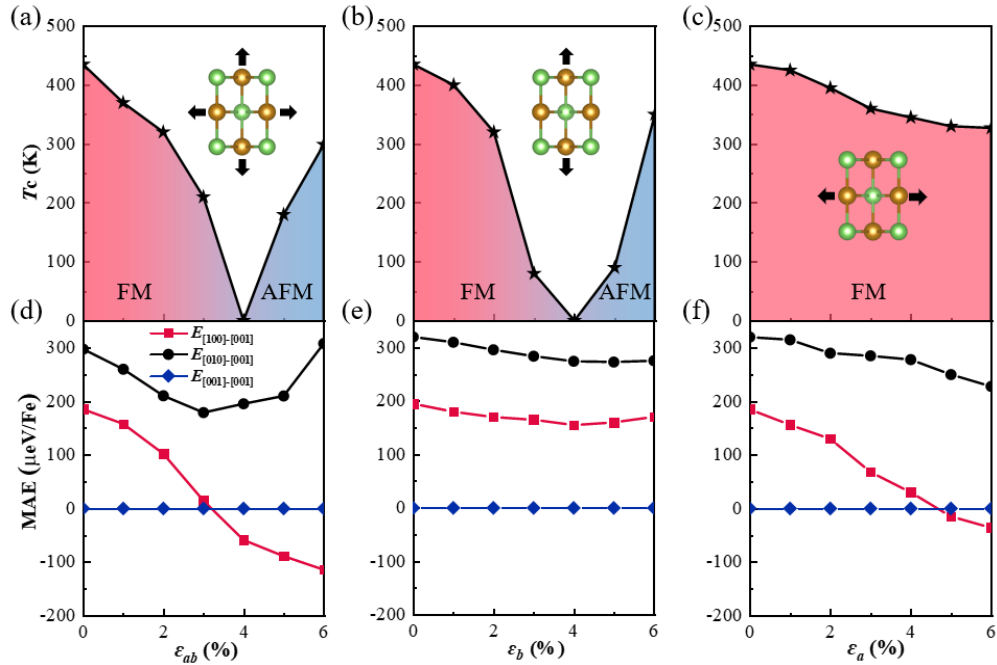

FIG. 6. Magnetic properties versus in-plane tensile strain ($\varepsilon$) in the range from 0% to 6%. (a-c) $T_c$, and (d-f) MAE. In (d)-(f), a negative MAE implies an easy axis along the in-plane uniaxial [100] direction.

Indeed, our calculations demonstrate that the FEL strain can easily modulate the $T_c$. As depicted in Fig. 6 (a) and (b), $T_c$ decreases rapidly with increasing biaxial tensile strain and uniaxial tensile strain applied along the long axial direction. Beyond ~4%, the $T_c$ reaches zero, meaning that the magnetic behavior in t-FeAs is transferred from FM to AFM. In contrast, the $T_c$'s are slightly changed in this 2D crystal under tensile strain applied along $a$ direction [Fig. 6(c)]. The strain-tunable behavior of magnetic coupling is crucial for practical electromechanical nanodevice applications. For example, the t-FeAs can be tailored to represent logic state as either "1" with FM state or "0" with AFM state, enabling the attainment of a spin switch. On the other hand, we find that the MAE change caused by strain is tunable. A moderate tensile strain for both the biaxial tensile strain and uniaxial tensile strain applied along the short axial direction will change the orientation of easy axis from out-of-plane into in-plane uniaxial direction [Fig. 6(d)-(f)]. The robust uniaxial magnetic anisotropy, mainly benefiting from its strain unbroken structural anisotropy, maintains the long-range ordered low temperature phase.

**B. Effect of U on magnetism of FeAs.**

Figure 7 shows the calculated total magnetic moment per Fe atom of t-FeAs using the GGA + $U$ method with $U_{eff}$ varied from 0 to 5 eV. The magnetic moment increases sharply with increasing $U_{eff}$ until it reaches 2.97 $\mu_B$ at $U_{eff}$ = 2 eV. This rather sensitive behavior is reasonable and expected as increasing $U_{eff}$ enhances the electron localization. Beyond this critical point, if we continue to increase $U_{eff}$, the curve becomes flat and the magnetic moment tends to get saturated, with the saturated magnitude reaching about 3.08 $\mu_B$. Consistent with the GGA + $U$ ($U_{eff}$ = 2 ~ 5 eV), the HSE06 functional shows a magnetic moment of 3.13 $\mu_B$. In addition, we have verified that our qualitative results are reasonable for $U_{eff}$ = 2 eV ~ 5 eV. Therefore, $U_{eff}$ = 3 eV is used throughout the paper.

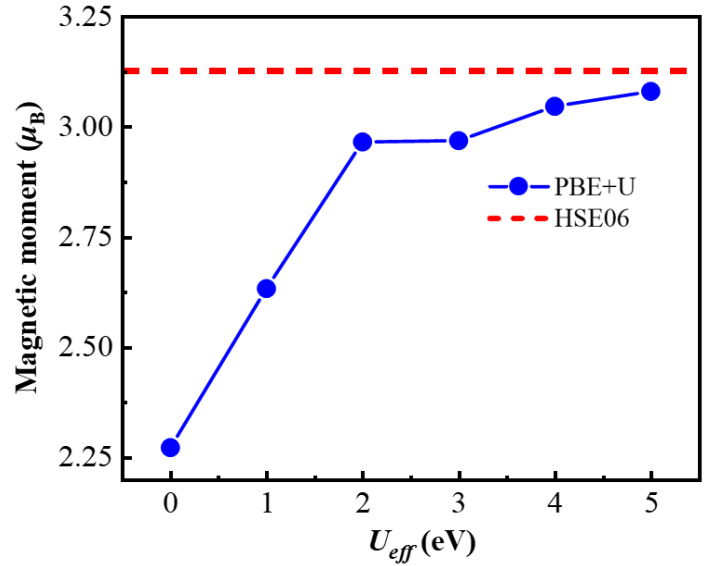

FIG. 7. Magnetic moment per Fe atom of t-FeAs calculated with the HSE06 (dashed line) and GGA + $U$ (solid line) functionals, where $U_{eff}$ is varied from 0 to 5 eV.

**C. Discussion about our results**

Before concluding, it is necessary to point out the following: (*i*) We highlight, as far as we know, that a tunable structural chiral-like atom deviation-nondeviation phase transition is discovered in a 2D system for the first time. It is also the first time to realize a strong magnetoelastic coupling in a FM/FEL multiferroic with robust uniaxial magnetic anisotropy. These two orders are significantly linked to ferroelastic phase transition, and could be directly probed by STM. (*ii*) Things will be different if we choose the NEB method to search the switching pathway between the initial and final state. Although an energy profile similar to the normal phase can be observed, a twist one and hence a structural phase transition often goes unrecognized. This indicates that the NEB method acts as an upper limit, as unknown pathways may exist with structural forms of even lower energy. (*iii*) Our four-state mapping analysis shows that MR-FeAs exhibits AFM coupling, consistent with ref. [42]. The remaining 3 polymorphs of monolayer FeAs are found not to exhibit FEL. As previously reported, RuSb, FeSb, CrN, VP, and VAs have the same crystal structure and internal atomic positions as t-FeAs, which are all FM-FEL multiferroic [25,27,67]. Our test calculations show that no analogous elastic strain-induced structural phase transition is observed (Fig. S9), indicating the role of composition in driving strain-induced structural phase transition. (*iv*) Although our work provides a stable multifunctional 2D material, it is metallic. Atomic monolayer magnetic materials with unique band structure, such as semiconductivity, half-metallicity, and topological Dirac states [67-70], are usually preferred for future high-performance multifunctional devices such as tunneling magnetoresistance sensors, magnetic random-access memories, and quantum computation/communication. The hunt for such 2D materials is still on.

## VII. CONCLUSIONS

On the basis of first-principles calculations and Monte Carlo simulation, we present a new 2D t-FeAs monolayer with rectangular lattice and large perpendicular magnetic anisotropy. Both phonon spectrum calculation and AIMD simulation indicate that t-FeAs is dynamically and thermally stable. We point out that the four magnetic configurations method has obvious limits in deriving the magnetic exchange parameters and magnetic ground state for a 2D material that possesses rectangular spin lattice. By employing the four-state mapping analysis, the proposed t-FeAs is identified as an intrinsic ferromagnet with high Curie temperature over 435 K. Both the $T_c$ and easy axis of t-FeAs are strain tunable. More importantly, we report for the first time a strain-mediated chiral-like atomic distortions behavior. Detailed analysis on the structure and magnetization let us propose that their coupling based on FEL switching could be directly probed, such as by STM. These highly nontrivial properties endow t-FeAs great potential for applications in sensing and information storage devices, and provide opportunities and a platform to explore unprecedented coupling physics beyond the current ones in multiferroic materials.

## ACKNOWLEDGMENTS

This work is partially supported by Fundamental Research Funds for the Central Universities (No. SWU-KR22030), by the National Natural Science Foundation of China (No. 12104294), by the Basic Scientific Research Expenses Program of Universities directly under Inner Mongolia Autonomous Region (No. JY20220213), by Research Program of Science and Technology at Universities of Inner Mongolia Autonomous Region (No. NJZY22389), and by Shanxi Supercomputing Center of China.

Competing Interests: Authors declare no competing interests.